# Chandrayaan-3 Alternate Landing Site: Pre-Landing Characterisation

K. Durga Prasad[1*], Dibyendu Misra [1,2], Amitabh[3], Megha Bhatt[1], G. Ambily[1,4], Sachana Sathyan [1,5], Neeraj Srivastava[1] and Anil Bhardwaj[1]

[1]Physical Research Laboratory, Ahmedabad-380009, India; [2] Indian Institute of Technology Gandhinagar, Gandhinagar-382055, India; [3]Space Applications Centre(ISRO), Ahmedabad-380015, India; [4] Andhra University, Visakhapatnam-530003, India; [5] University of Kerala, Trivandrum-695581, India

***Corresponding Author Email:** durgaprasad@prl.res.in

**Abstract:**

India's third Moon mission Chandrayaan-3 will deploy a lander and a rover at a high latitude location of the Moon enabling us to carry out first ever in-situ science investigations of such a pristine location that will potentially improve our understanding on primary crust formation and subsequent modification processes. The primary landing site (PLS), is situated at 69.367621 °S, 32.348126 °E. As a contingency, an alternate landing site (ALS) was also selected at nearly the same latitude but ~450 km west to PLS. In this work, a detailed study of the geomorphology, composition, and temperature characteristics of ALS has been carried out using the best-ever high resolution Chandrayaan-2 OHRC DEMs and Ortho-images, datasets obtained from Chandrayaan-1 and on-going Lunar Reconnaissance Orbiter. For understanding the thermophysical behaviour, we used a well-established thermophysical model. We found that the Chandrayaan-3 ALS is characterised by a smooth topography with a relatively elevated central part. ALS is dominated by ejecta of the Moretus-A crater of Eratosthenian age and is situated on the ejecta blanket of Tycho crater. The ALS is a scientifically interesting site with a high possibility of sampling ejecta materials from Tycho and Moretus. However, due to presence of Eratosthenian age ejecta materials, the site is boulder rich, The OHRC derive hazard map confirms 75% of hazard-free areas within ALS and thus suitable for landing and rover operations. Sampling traces of Tycho ejecta with APXS and LIBS onboard rover will be useful in understanding compositional variations within ALS. Based on the spectral and elemental analysis of the site, Fe is found to be ~ 4.8 weight percent (wt.%), with Mg ~ 5 wt.%, and Ca ~ 11 wt.%. Compositionally, ALS is similar to PLS with typical highland soil type composition. Spatial and diurnal variability of ~40 K and ~175 K has been observed in the surface temperatures at ALS. Although belonging to similar location like PLS, ALS showed reduced day-time temperatures and enhanced night-time temperatures compared to PLS, indicating a terrain of distinctive thermophysical characteristics compared to that of PLS. Like PLS, ALS is also seems to be an interesting site for science investigations and Chandrayaan-3 is expected to provide new insights into the understanding of lunar science even if it happens to land in the alternate landing site.

# 1. Introduction

ISRO's third mission to the Moon, Chandrayaan-3 has been launched on 14 July 2023 and is currently in an orbit around the Moon to make a soft landing on 23 August 2023. Chandrayaan-3 consist of a propulsion module, lander and rover accommodating six payloads for carrying out scientific investigations in the vicinity of the landing site[1-6]. Chandrayaan-3 lander along with its rover is intended to land at a high-latitude region of the Moon for which two sites – primary (PLS) and alternate (ALS), have been finally identified adhering to both technological constraints and scientific merits. The primary landing site is situated at 69.367621 ˚S, 32.348126˚E and found to be safe for landing with slope less than 4˚ in about 78 per cent of landing area[7] To handle any contingency prior to landing, an alternate site is also identified at 69.497764 ˚S, 17.330409 ˚W[7] which may be encountered after 3 to 4 days of the nominal prime landing (Private communication). The locations of the PLS and ALS are shown in Figure 1. A detailed contextual characterisation of PLS has been carried out in terms of geomorphological, compositional and thermophysical perspectives recently by 7. In a similar line, we conducted a detailed characterisation study of the alternate landing site for planning mission operations and interpretation of science data obtained from onboard instruments, in case it happens to be the final site of landing. Variability within the local terrain, illumination and surface temperatures were studied to assist for safe landing, carryout lander operations and also for rover path planning. For interpretation of observed in-situ data and optimal science derivation, a detailed geomorphological, compositional and thermophysical characterisation is the primary objective of carrying out this study. These characterisation studies are based on all relevant datasets available from previous orbiter missions. The ALS is confined to ~2.4 x 4 km and its geomorphological characterisation is based on Chandrayaan-3 specific targeted observations from Chandrayan-2 orbiter, particularly the best spatial resolution (25 cm) images from OHRC (Orbiter High Resolution Camera)[8] and derived digital elevation model (DEM). While we focus only on understanding the characteristics of a smaller region of 4 km x 2.4 km selected for landing, a larger perspective can be obtained based on the studies similar to that reported in 9.

# 2. Datasets and methodology

High-resolution DEM derived from OHRC with a spatial resolution of ~25 cm, is used for the geomorphological studies and thermophysical modelling[10,11]. Various other datasets from Chandrayaan-1, SELENE and LRO have also been used for the study. Spectral/compositional studies have been carried out using the Moon Mineralogy Mapper ($M^3$) of Chandrayaan-1[12].

Topography is obtained from LOLA and WAC images[13] and surface brightness temperatures are derived from Diviner onboard LRO[14]. PRL 3D Thermophysical Model is used to carry out the modelling of the site in local scale[15].

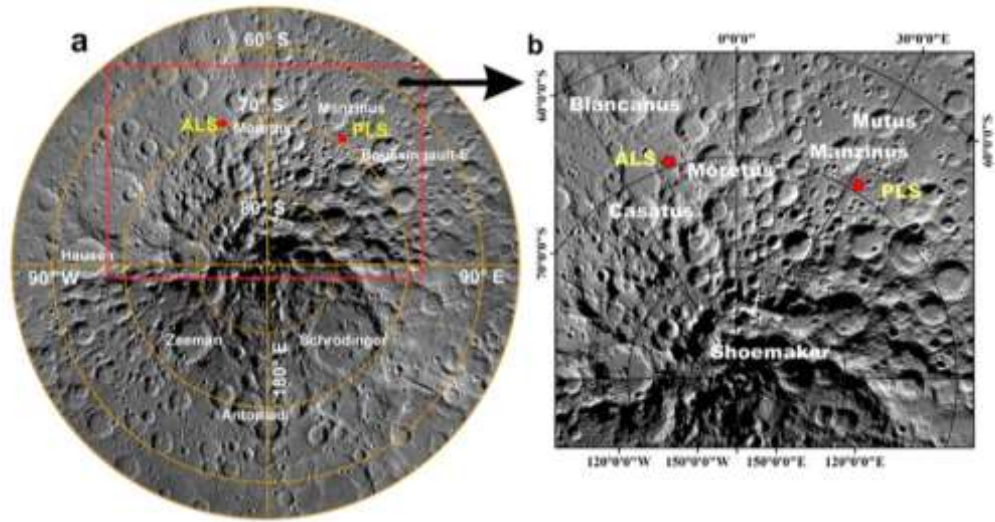

**Figure 1:** (a) Locations of the primary (PLS) and alternate (ALS) landing sites were plotted over the LRO-WAC mosaic using the lunar south polar projection. The area near these landing sites (marked in red box) was represented in the following image (b) Perspective view of ALS with respect to PLS and lunar south pole.

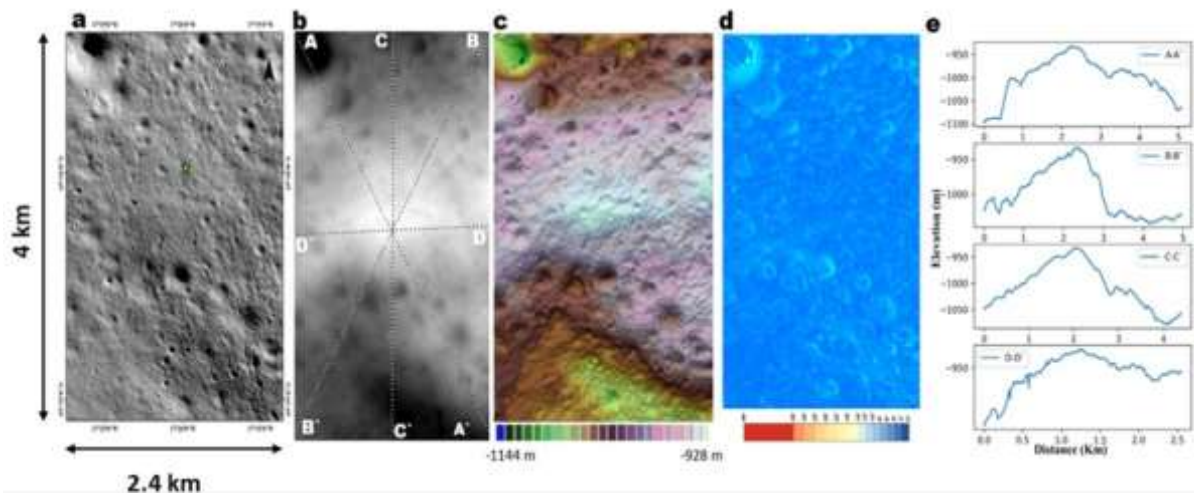

**Figure 2:** Detail geomorphic maps (a-e) of the alternate landing site (ALS). (a) OHRC image (equidistance cylindrical) where ALS is marked in green colored star, (b) OHRC derived DEM, (c) OHRC derived colour-coded DEM, (d) Slope map, and (e) Elevation profiles through A-A`, B-B`, C-C`, and D-D` respectively.

## 3. Geomorphological study

Fig. 1 shows the locations of identified PLS and ALS for Chandrayaan-3 landing. The ALS and PLS are both situated at similar latitudes (~69°S) but about ~450 km apart. Initially Terrain Mapping Camera (TMC) images of Chandrayaan and derived digital elevation model (DEM),

SELENE and LOLA derived DEMs and LRO WAC and NAC images were used to find a suitable alternate site in western longitudes for the same latitude as Primary Landing Site[11]. A suitable site of 4 km x 2.4 km was selected as ALS [69.497764° S, 17.330409° W], ~ 450 km west of the primary landing site (PLS) [69.36762° S, 32.348126° E] of the Chandrayaan-3 as shown in Fig. 1[7] . In this work, we used highest resolution OHRC derived DEM for understanding the landing site terrain in detail and based on OHRC images, hazard map is prepared similar to PLS[7] .

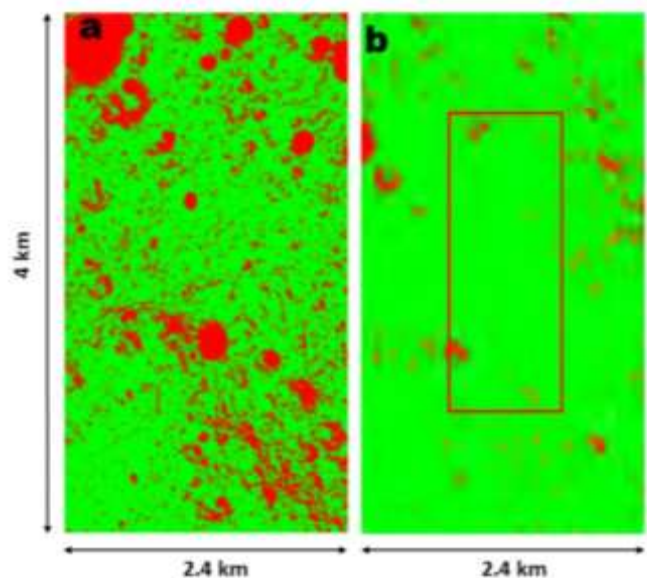

**Figure 3:** OHRC derived (a) hazard map and (b) shadow map over the alternate landing site where safe areas were marked in green and hazardous areas in red color.

The ALS OHRC image, and derived parameters from DEM are shown in Fig. 2. We found that the ALS belong to smooth topography with a comparatively elevated central part (Fig. 2.c, and e). The average elevation variation is 216 m within the area with an overall slope of less than 7° (Fig. 2.d), which satisfied the criteria used for the selection of PLS of Chandrayaan-3[7] . The ALS is situated on the west of the Moretus crater (Fig. 1) which belongs to the crater cluster unit of the Eratosthenian age[16]. Hazard maps based on slope and illumination were derived in the same way as detailed in (7). The hazard map shown in Fig. 3 cover nearly 75% of hazard-free areas suggesting that the ALS is suitable for landing and rover operations. Fig. 4 shows a geomorphology map of ALS. We found two distinct geomorphic units; a and b as marked in Fig. 4 based on the surface textural variation within ALS. Interestingly, both the geomorphic units belong to the comparable slope variations (Fig. 2d) but are distinguishable mainly based number density of fresh craters in vicinity. The surficial texture variations shown at OHRC scale within such a small area could be mainly due to the ejecta of the Moretus-A crater which covers the region shown in Fig. 4. On the south-western (SW) flank of the ALS, several fresh craters are observed with high density of boulders as shown in Fig. 4. As shown in Fig. 4, ALS is found to be interesting from geomorphology perspective. In order to understand the albedo variations within ALS, we used the Lunar orbiter laser altimeter (LOLA) data measured using laser pulse at 1064 nm. The LOLA provides the relative reflectivity of the surface at zero phase angle[17]. Areas with relatively high albedo indicate fresh ejecta or immature craters. The albedo variation of LOLA shown in Fig. 5 suggest presence of ejecta materials at ALS. Tycho crater (43.37° S, 11.35° W) is situated north of ALS which is one of the youngest craters of the Moon with (~100 Ma) with prominent visible ejecta rays[18]. The ALS with overlapped ejecta traced back to crater Tycho might provide a unique opportunity to sample Tycho crater composition at ALS and the results can also be cross-validated and compared to Apollo 17 landing site[19].

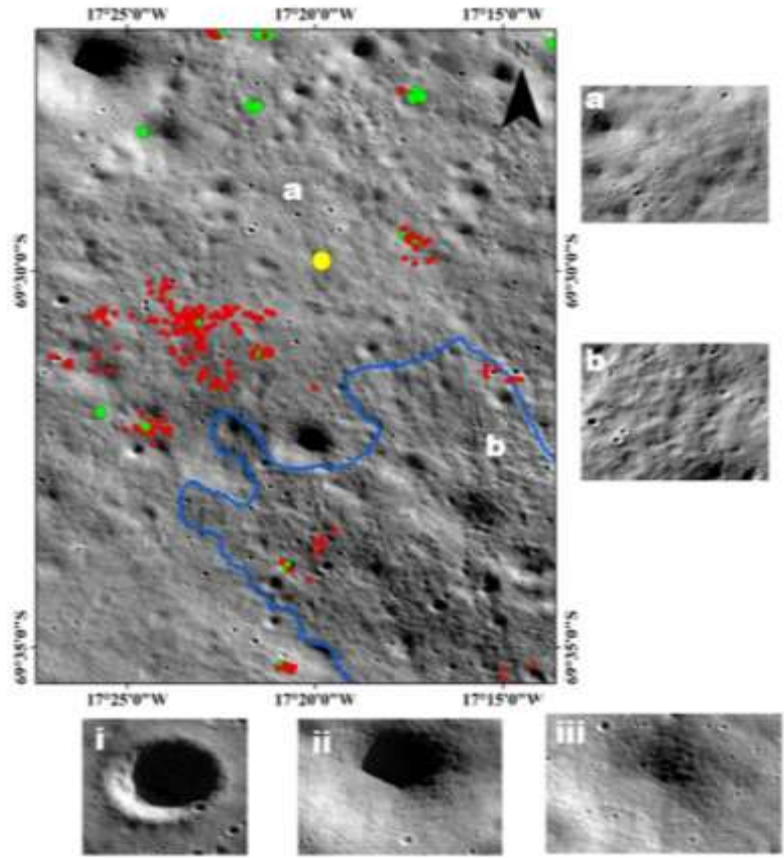

**Figure 4:** Geomorphological map around the ALS (marked in Yellow) of the Chandrayaan-3. The entire area was divided into two geomorphic units (a and b) by the blue dashed line. (a) Unit 'a' was relatively smooth whereas (b) unit 'b' had a rough surface with relatively low albedo. Fresh craters (green) and boulders (red) were marked as the potential sampling sites over the region. Three types of craters (i-iii) were noticed around the ALS. (i) On the SW of the ALS, a fresh crater with boulders, and on the NW (ii) degraded, and (iii) Ghost crater were encountered.

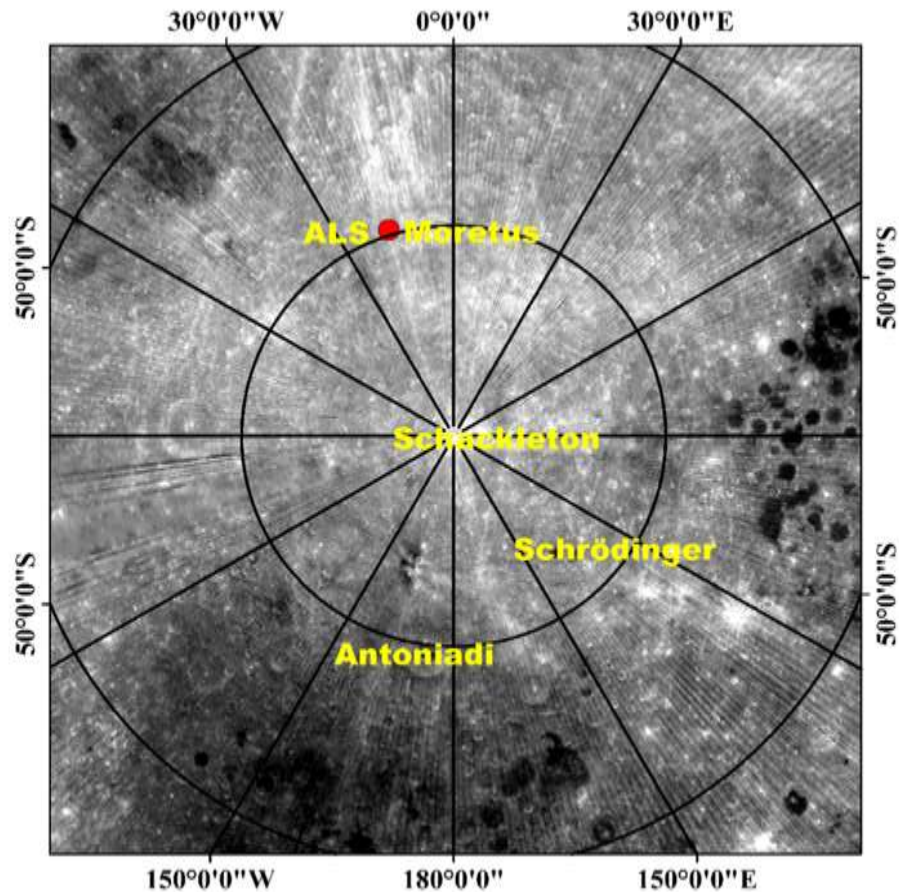

**Figure 5:** LOLA albedo map was projected in the lunar south polar projection where the ALS (red-colored point) was plotted east to the Moretus crater. Higher reflectivity was represented as white color. It can be observed that the area near the ALS is optically immature, which might have occurred due to the superposition of the fresh ejecta from the Tycho crater.

## 4. Mineralogical and compositional study

Thermally and photometrically calibrated, level 2 $M^3$ reflectance (ID: M3G20090206T185403) data of optical period 1B[20] covers the ALS. We used $M^3$ data to understand the mineralogical and compositional variations within ALS. Typical spectral characteristics of ALS is shown in Fig. 6. We found that the $M^3$ reflectance spectra do not show any prominent absorption band around 1 and 2 μm (Fig. 6b). These are typical highland type reflectance spectra associated with shocked plagioclase lithology since plagioclase loses its crystal structure (an absorption feature around 1.25 μm) relatively faster due to meteoritic impacts[21,22]. We observed an increase in reddening in the reflectance spectra belong to ALS and that could be a result of space weathering[23,24]. Through spectral data, we could not identify Tycho ejecta distinct from its surroundings. Another typical spectral characteristic from this region is confined to a few fresh craters that show a weak absorption features around 1 μm with relatively high reflectance (Fig. 6b). However, this band do not show a characteristic shape of Olivine. Such a spectral characteristic could be due to plagioclase-dominated pyroxene (∼1 wt.%) mixed soils[25]. These regions will be interesting to explore through rover as they could be resulted from fresh crater ejecta[26]. The absorption band depth of the absorption feature around 1 μm is below 5% and difficult to characterize due to lower Signal to Noise ratio (SNR). We observed a consistent dip around 2.8 μm in most of the M3 reflectance spectra from ALS. The $M^3$ data used here belongs to morning hour's observations and the absorption feature around 2.8 μm is most likely due to solar wind interaction representing unstable water component within ALS.

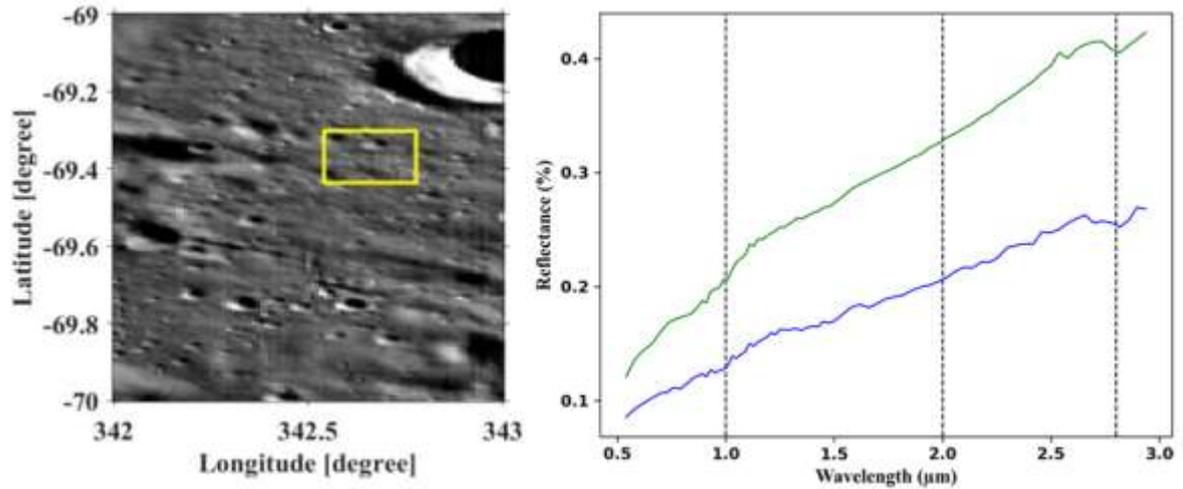

**Figure 6:** (a) 1˚ x 1˚ M3 albedo image including Chandrayaan-3 ALS (marked as a yellow box) represented at ~1500 nm. (b) Extracted representative spectra from ALS are plotted. The dashed black lines are given at 1, 2, and 2.8 μm for reference. Highland-type soil is represented through the blue-colored spectra that have no significant absorption feature at 1 μm and 2 μm. Spectra colored in green extracted from some fresh craters where a minor absorption at 1 μm can be observed. The OH/H2O spectral signature around 2.8 μm is also detected in both the representative spectra.

For understanding average chemical composition of ALS, we extracted average abundances of Fe, Mg, and Ca from the $M^3$ based global elemental abundance maps at 1.5 km/pixel spatial resolution[27]. The spectral analysis outcomes are in agreement with the elemental abundances extracted. Fe is found to be ~ 4.8 weight percent (wt.%), with Mg ~ 5 wt.%, and Ca ~ 11 wt.%. The average elemental composition of ALS suggests that the ALS of Chandrayaan-3 has an anorthositic highland composition that has experienced extended space weathering throughout the lunar geological time. The ALS is found to be similar to PLS compositionally[7].

## 5. Temperatures and Thermophysical characterisation

Understanding the surface temperatures and their variability in the vicinity is not only necessary for estimating the survivability and lifetime of the lander and the rover but also an important parameter for to direct their operations. This will also help in understanding and interpreting the data received and deriving science from Chandra's Surface Thermophysical Experiment (ChaSTE) and Instrument for Lunar Seismic activity (ILSA) instruments onboard the Chandrayaan-3 lander. Therefore, a detailed thermophysical characterisation of the alternate landing site has been carried out to understand the temperatures and thermal behaviour in the vicinity both at regional as well as local scales. This thermophysical study has been carried out utilising datasets from Diviner radiometer[14] onboard LRO and a three-dimensional thermophysical model[15]. Diviner RDR datasets are processed to derive the diurnal brightness temperatures observed at the 4.2 x 2.4 km area. Processing methodology of Diviner datasets used for this study is detailed in 7. The optimised data is then overlaid over LOLA DEM to interpret the temperature variations in 3-dimension. Figure 7 depicts the Diviner derived temperature variability map at local noon of the alternate landing site. It is evident that even at local scales, surface temperatures seem to exhibit a significant spatial variation, possibly due to small-scale topographical differences, which is an interesting aspect. The spatial variability of local surface temperatures within ALS region was derived for various phases of the day – dawn, noon, dusk and midnight. However, surface temperatures corresponding to local noon is only shown in figure 7(a) as maximum variation in temperatures is expected only during this time. Being a very small area (4 km x 2.4 km), significant temperature variability is not expected within. However, a variation of ~40K variation within the surface temperatures of the region is seen as evident in figure 7(a) which is usually not expected as the landing area is relatively small (~2.4 km x 4 km). Figure 7(b) shows the diurnal variability of minimum and maximum surface temperatures for ALS which is as expected[28,29,15]. From Diviner observations, surface temperature variation of ~60 K – 270 K was observed. While the trend of temperature variability seems to be similar to that of PLS[7], the recorded maximum temperatures are relatively lower and minimum temperatures are slightly higher directing towards local surface of thermophysical behaviour quite distinct from that of PLS.

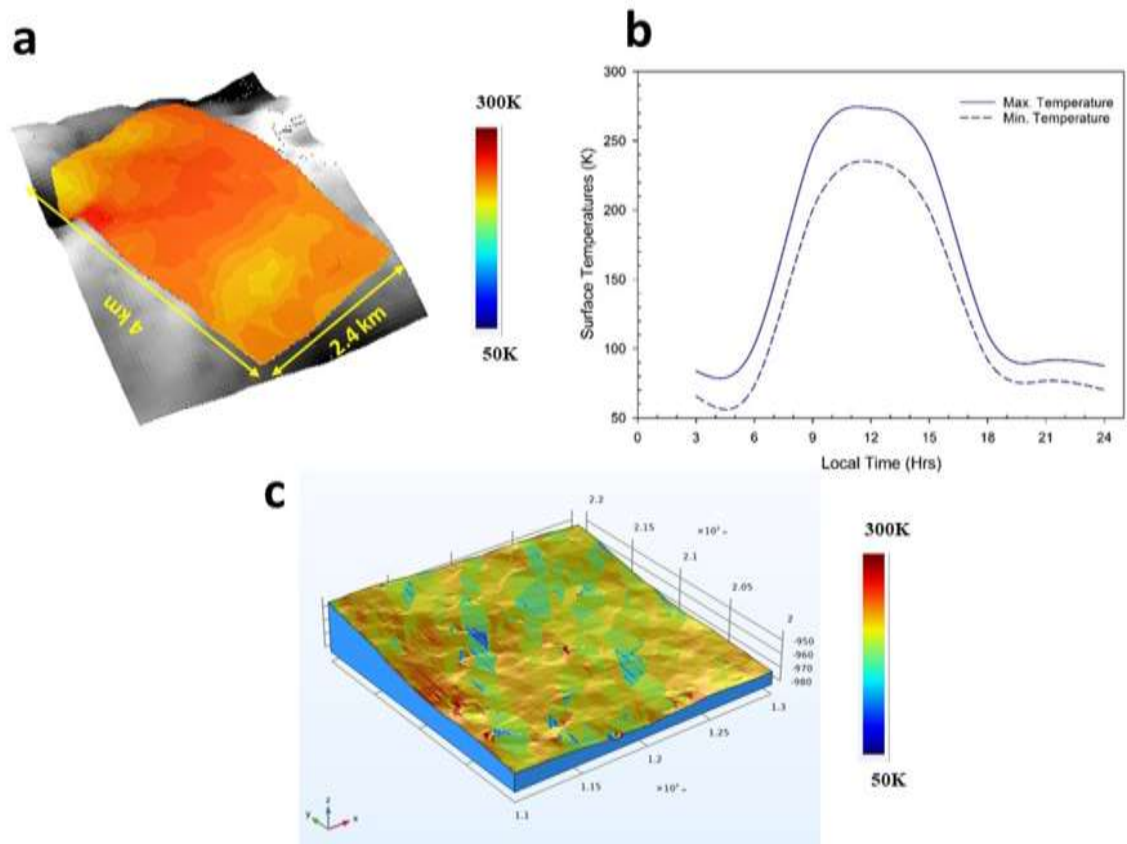

**Figure 7:** (a) Local noon surface temperature variations for the landing region of the alternate site (b) Variability of minimum and maximum surface temperatures at ALS landing region (c) Model derived surface temperature variation within 200m x 200m area at the centre of the ALS during dawn phase, depicting distinct thermophysical behaviour at local scale. The patches seen in figure 7(c) are interpolation artefacts and may be ignored.

Using a three-dimensional model[15], the thermophysical behaviour at local scale for an area of 200 m x 200 m has been computed in order to understand the surface and subsurface temperatures that are likely to be encountered by ChaSTE experiment onboard the Chandrayaan-3 lander. High resolution topography data from OHRC has been used for model simulations. The model also includes the effect of an outermost insulating layer of ~9 cm thick corresponding to the estimations given by 28.The boundaries are thermally insulated and solar irradiation corresponding to the landing site is given in x,y,z directions. Thermophysical parameters such as effective thermal conductivity, specific heat are given as analytic functions. The model has run by finite element method, where the geometry is meshed into finer surface elements, with size varying from 11m to 0.8m. The model has run for 90 earth days and seemed to be equilibrated by the run time. The results exhibited distinct thermal variations at the site. Considering the relative smooth surface at ALS, the temperatures seem to be surprisingly different even within the 200m x 200m area. The model-derived surface temperature for a small region of 200 m x 200 m around the centre of ALS during dawn phase is shown in figure 7(c). Distinct variability of local scale surface and subsurface temperatures are evident from model simulations that could possibly either due to small topographic variations or surface of distinct thermophysical characteristics or both.

## 5. Conclusion

Chandrayaan-3 is India's third mission to the Moon which will deploy a lander and a rover at a high latitude location of the Moon for carrying out first ever in-situ science investigations. After an in-depth evaluation of several sites, two sites – a primary site and an alternate site for contingency, have been selected[11].. We have carried out a detailed characterisation of both the landing sites. While the primary landing site study is reported in[7], details about alternate site are reported in this work. Detailed geomorphological, compositional and thermophysical characterisation of the alternate landing site has been carried out using the best-ever high resolution OHRC DEMs and Ortho-images, other datasets and modelling. Our hazard map using OHRC suggest that ~75% of the landing area is hazard-free and safe to land. Local geomorphological variations show that ALS is characterised by a smooth topography with a relatively elevated central part with an average elevation of 216 m. The ALS, which lies to the west of the Moretus crater, can be divided into two distinct geomorphic units primarily based on fresh crater distribution and associated boulder density. The ALS also received ejecta from crater Tycho which might provide an opportunity to sample Tycho crater composition. Based on the spectral analysis of the site, Fe is found to be ~ 4.8 (wt.%), with Mg ~5 wt.%, and Ca ~11 wt.%, suggesting typical highland composition which is similar to PLS. The temperature and thermophysical analysis of ALS showed a significant spatial and diurnal variability of surface temperatures as seen in the case of PLS. However, a relatively lower day-time and higher night-time temperatures in comparison with that of the PLS suggests that the ALS may have different thermophysical parameters than the PLS. Thus, along with the PLS, the ALS of Chandrayaan-3 is also expected to provide new insights into the understanding of lunar science.

**Acknowledgements:** We thank the Department of Space, Govt. of India for providing the financial support for carrying out this work. We thank Department of Science and Technology (DST), Government of India, for supporting G. Ambily & Sachana Sathyan under INSPIRE PhD Fellowship. A. Bhardwaj was supported by J. C. Bose Fellowship during the period of this work. We thank Shri. Nilesh M Desai, Director, Space Applications Centre(SAC), Ahmedabad and Shri Debajyoti Dhar, DD, SIPA, SAC, Ahmedabad, for providing necessary support for carrying out data analysis of OHRC. We thank Chandrayaan-2, 3 mission, project teams and ISSDC for their data support. The local scale model computations were performed on the PARAM Vikram-1000 High Performance Computing Cluster of the Physical Research Laboratory, Ahmedabad.

**References:**

1. Bhardwaj A. , 2021, In 43rd COSPAR Scientific Assembly during Jan. 28 - Feb. 4,in Sydney, Australia (virtually), Vol. 43, p. 765, published by COSPAR
2. Manju G. , Pant, T.K., Sreelatha, P., Nalluveettil, S.J., Kumar, P .P ., Upadhyay, N.K., Hossain, M.M., Naik, N., Yadav, V.K., John, R. and Sajeev, R., 2020. Lunar near surface plasma environment from Chandrayaan-2 Lander platform: RAMBHA-LP payload.,2020,CURRENT SCIENCE,118,383.
3. Durga Prasad K. , 2016, Annual Report 2015 16-00, Front-End Electronics Development for ChaSTE P ayload onboard Chandrayaan-2 Lander. Physical Research Laboratory, Navrangpura, Ahmedabad
4. John, J., Thamarai, V., Mehra, M. M., Choudhary, T., Giridhar, M. S., Jambhalikar, A., .& Laxmiprasad, A. S. (2020). Instrument for Lunar Seismic Activity Studies on Chandrayaan-2 Lander. CURRENT SCIENCE, 118(3), 376.
5. Shanmugam, M., Vadawale, S. V., Patel, A. R., Mithun, N. P. S., Adalaja, H. K., Ladiya, T., ... & Acharya, Y. B. (2020). Alpha Particle X-ray Spectrometer onboard Chandrayaan-2 Rover. Current Science (00113891), 118(1).
6. Laxmiprasad A. S. , Raja V. S., Menon S., Goswami A., Rao M. V. H., Lohar K. A., 2013, Adv. Space Res. , 52, 332
7. Durga Prasad Karanam and others, Contextual Characterisation Study of Chandrayaan-3 Primary Landing Site, *Monthly Notices of the Royal Astronomical Society: Letters*, 2023; slad106, https://doi.org/10.1093/mnrasl/slad106
8. Chowdhury, A. R., Saxena, M., Kumar, A., Joshi, S. R., Amitabh, A. D., Mittal, M., ... & Gupta, A. (2019). Orbiter high resolution camera onboard Chandrayaan-2 orbiter. Current Science, 117(7), 560.
9. Sinha, R. K., Rani, A., Ruj, T., & Bhardwaj, A. (2023). Geologic investigation of lobate scarps in the vicinity of Chandrayaan-3 landing site in the southern high latitudes of the moon. Icarus, 402, 115636.
10. Amitabh et al. 2021, In 52nd Lunar and Planetary Science Conference, held 15-19 March, 2021 at The Woodlands, Texas and virtually. LPI Contribution No. 2548, High Resolution DEM Generation from Chandrayaan-2 Orbiter High Resolution Camera Images. p. 1396

11. Amitabh , Suresh K. , Prashar A. K., Suhail. 2023, In 54th Lunar and Planetary Science Conference, held 13-17 March, 2023 at The Woodlands, Texas and virtually. LPI Contribution No. 2806, id.1037, Terrain Characterisa- tion of Potential Landing Sites for Chandrayaan-3 Lander using Orbiter High Resolution Camera (OHRC) Images

12. Pieters, C. M., Boardman, J., Buratti, B., Chatterjee, A., Clark, R., Glavich, T., ... & White, M. (2009). The Moon mineralogy mapper (M³) on chandrayaan-1. Current Science, 500-505.

13. Smith, D. E., Zuber, M. T., Jackson, G. B., Cavanaugh, J. F., Neumann, G. A., Riris, H., ... & Zagwodzki, T. W. (2010). The lunar orbiter laser altimeter investigation on the lunar reconnaissance orbiter mission. Space science reviews, 150, 209-241.

14. Paige, D. A., Foote, M. C., Greenhagen, B. T., Schofield, J. T., Calcutt, S., Vasavada, A. R., ... & McCleese, D. J. (2010). The lunar reconnaissance orbiter diviner lunar radiometer experiment. Space Science Reviews, 150, 125-160.

15. Durga Prasad, K., Rai, V. K., & Murty, S. V. S. (2022). A Comprehensive 3D Thermophysical Model of the Lunar Surface. Earth and Space Science, 9(12), e2021EA001968.

16. Wilhelms, D.E., Howard, K.A. and Wilshire, H.G., 1979. Geologic map of the south side of the Moon. Department of the Interior, US Geological Survey.

17. Lemelin, M., Lucey, P. G., Neumann, G. A., Mazarico, E. M., Barker, M. K., Kakazu, A., ... & Zuber, M. T. (2016). Improved calibration of reflectance data from the LRO Lunar Orbiter Laser Altimeter (LOLA) and implications for space weathering. *Icarus*, *273*, 315-328.

18. Stöffler, D., & Ryder, G. (2001). Stratigraphy and isotope ages of lunar geologic units: Chronological standard for the inner solar system. *Space Science Reviews*, *96*(1-4), 9-54.

19. Lucchitta, B. K. (1977). Crater clusters and light mantle at the Apollo 17 site; a result of secondary impact from Tycho. *Icarus*, *30*(1), 80-96.

20. Green, R.O., Pieters, C., Mouroulis, P., Eastwood, M., Boardman, J., Glavich, T., Isaacson, P., Annadurai, M., Besse, S., Barr, D. and Buratti, B., 2011. The Moon Mineralogy Mapper (M3) imaging spectrometer for lunar science: Instrument description, calibration, on-orbit measurements, science data calibration and on-orbit validation. Journal of Geophysical Research: Planets, 116(E10).

21. Adams, J. B., Horz, F., & Gibbons, R. V. (1979, March). Effects of shock-loading on the reflectance spectra of plagioclase, pyroxene, and glass. In *LUNAR AND PLANETARY SCIENCE X, P. 1-3. Abstract.* (Vol. 10, pp. 1-3).

22. Yamamoto, S., Nakamura, R., Matsunaga, T., Ogawa, Y., Ishihara, Y., Morota, T., ... & Haruyama, J. (2015). Featureless spectra on the Moon as evidence of residual lunar primordial crust. *Journal of Geophysical Research: Planets*, *120*(12), 2190-2205.

23. Hapke, B., 2001. Space weathering from Mercury to the asteroid belt. Journal of Geophysical Research: Planets, 106(E5), pp.10039-10073.

24. Pieters, C.M. and Noble, S.K., 2016. Space weathering on airless bodies. Journal of Geophysical Research: Planets, 121(10), pp.1865-1884.

25. Nash, D. B., & Conel, J. E. (1974). Spectral reflectance systematics for mixtures of powdered hypersthene, labradorite, and ilmenite. Journal of Geophysical Research, 79(11), 1615-1621.

26. Bernhardt, H., Robinson, M. S., & Boyd, A. K. (2022). Geomorphic map and science target identification on the Shackleton-de Gerlache ridge. *Icarus*, *379*, 114963.

27. Bhatt, M., Wöhler, C., Grumpe, A., Hasebe, N. and Naito, M., 2019. Global mapping of lunar refractory elements: Multivariate regression vs. machine learning. Astronomy & Astrophysics, 627, p. A155.

28. Hayne, P. O., Bandfield, J. L., Siegler, M. A., Vasavada, A. R., Ghent, R. R., Williams, J. P., ... & Paige, D. A. (2017). Global regolith thermophysical properties of the Moon from the Diviner Lunar Radiometer Experiment. Journal of Geophysical Research: Planets, 122(12), 2371-2400.

29. Vasavada, A. R., Paige, D. A., & Wood, S. E. (1999). Near-surface temperatures on Mercury and the Moon and the stability of polar ice deposits. *Icarus*, *141*(2), 179-193.